\documentclass[pra,showpacs,aps,amsmath,amssymb]{revtex4}
\usepackage{graphics}
\usepackage{amsmath,amsfonts,amssymb}
\usepackage{bm}
\usepackage{epsfig}
\usepackage[english]{babel}
\usepackage{xcolor}
\usepackage{ulem}

\begin{document}


\title{Comparative analysis of the Compton ionization of Hydrogen and Positronium. }

\author{Ivan S. Stepantsov$^1$} 
\email{i.stepantsov33@gmail.com (corresponding author)}
\author{Igor P. Volobuev$^2$, Yuri V. Popov$^{2,3}$}

\address{
$^1$ Physics Faculty, Lomonosov Moscow State University, Moscow, Russia\\
$^2$Skobeltsyn Institute of Nuclear Physics, Lomonosov Moscow State University, Moscow, Russia\\
$^3$Joint Institute for Nuclear Research, Dubna, Russia\\
}

\begin{abstract}

The paper deals with the Compton disintegration of positronium and a comparison of the differential cross sections of this process with the similar cross sections in the case of the Compton ionization of the hydrogen atom. Special attention is paid to the resonances arising, when the electron and positron move parallel to each other in continuum states with the same velocities and zero relative momentum. It is likely that a manifestation of this effect in the double differential cross section is found as an additional peak, which grows with a decrease in the photon energy.

\end{abstract}


\maketitle

\Large

\section*{1. INTRODUCTION}

Traditionally, we believe that the hydrogen atom is the simplest
quantum system, about which everything is known and which is used
as a basic target for the study of various atomic reactions:
excitation, ionization, capture, photoprocesses, etc. It is really
so, but there is one more simple quantum object with the same
properties, this is the positronium (the so-called exotic atom). The
singlet parapositronium state lives 0.12 ns,  the triplet
orthopositronium state lives 138.6 ns. These values can be
calculated theoretically \cite {AB}. A hydrogen atom lives on
average 0.1 - 10 ns in an excited state, although highly excited
states  can exist in vacuum for up to several seconds. In this
sense the lifetime of different forms of positronium is quite
comparable with the lifetime of an excited hydrogen atom, which
means that these atoms can be experimentally studied
with the same devices. Beyond the lifetime, the main difference
between hydrogen and positronium is the mass of the positively
charged particles: the proton mass is $ M = 1836.4 $ a.u., whereas
the mass of the positron (equal to that of the electron) is $ m =
1 $ a.u., and this difference results in significant differences
in the scattering cross sections.

In paper \cite{EPJD20}, the Compton ionization of a hydrogen atom
by a high-energy photon of several keV was discussed in detail,
and interesting features of various differential cross sections
were noted. This paper was inspired by the latest experiments on
the Compton ionization of helium atoms using the COLTRIMS detector
\cite{Nature20}. A characteristic feature of these experiments is
the measurement in coincidence of the momenta of an electron and
the residual ion in the range of angles practically equal to $
4\pi $, whereas the angle and energy of the scattered photons are
calculated from the conservation laws. The motivation behind the
paper was the possibility to use such a reaction for the purpose
of direct studies of the momentum distribution of an active
electron in an atom \cite{JQSRT}.

The experiments on Compton scattering at atoms without measuring
the final photon are pioneering. Back in the early 1920s,
following the discovery of the Compton effect, W. Bothe proposed
to use the Compton ionization of a bound electron by a photon to
study its momentum distribution in the target \cite{Bothe}. For
this purpose, he developed and implemented a scheme for measuring
in coincidence the momenta of the final photon and electron. This
scheme was imperfect, the accuracy was low, and mainly the photons
scattered forward were measured. Therefore, predominantly solid
targets were used. Quite different possibilities for working with
atoms and molecules were opened by the COLTRIMS detector, which
makes the work with a cold positronium beam not a fantasy.

In a sense, this paper is a continuation of paper \cite{EPJD20},
and here we will compare some cross sections of the Compton
ionization of the hydrogen atom (H) and positronium (Ps). In
particular, we will discuss in detail an interesting effect that
occurs, when, after the positronium decay, the electron-positron
pair moves with zero relative momentum.

We will use the atomic system of units: $ m_e = \hbar = | e | = 1
$. In these units, the speed of light is $ c = 137 $ and the fine
structure constant $ \alpha = 1/c $.

\section*{2. THEORY}

The non-relativistic Hamiltonian of a positronium atom in the
electromagnetic field of a laser has the following form (particle
1 is an electron, particle 2 is a positron):
$$
H = \frac{1}{2}\left(-i\vec\nabla_{r_1} - \frac{1}{c}\vec A(\vec r_1,t)\right)^2+\frac{1}{2}\left(-i\vec\nabla_{r_2} + \frac{1}{c}\vec A(\vec r_2,t)\right)^2+V(\vec r_1-\vec r_2) \eqno(1)
$$
In (1) $V=-1/|\vec r_1-\vec r_2|$  is the Coulomb potential of the
interaction between an electron and a positron, $\vec A$ is the
vector potential of the electromagnetic field. The Coulomb gauge
$\vec\nabla_r\cdot \vec A(\vec r,t)=0$ is used

We put
$$
\frac1c\vec A(\vec r,t) = \sqrt{\frac{2\pi}{\omega_i}}\ \vec e_i\ e^{i(\vec k_i\vec r-\omega_i t) } + \sqrt{\frac{2\pi}{\omega_f}}\ \vec e_f\ e^{-i(\vec k_f\vec r-\omega_f t) } + (c.c.). \eqno (2)
$$
In Eq. (2) $\vec e_i (\vec e_f)$ are the linear polarizations of
the initial (final) photons, $\vec k_i (\vec k_f)$ denote their
momenta, and the frequency (energy) of the photon is $\omega =kc$.
Since $(\vec k_i\cdot\vec e_i)=(\vec k_f\cdot\vec e_f)=0$, the chosen Coulomb gauge is
obviously fulfilled. This choice of the vector potential
corresponds to one absorbed and one emitted photon. Let us write
down the potential of interaction of an electron and a photon
$$
V_{int}^{el}= i\frac1c\ (\vec A(\vec r_1,t)\cdot\vec\nabla_1) +\frac{1}{2c^2}A^2(\vec r_1,t)\ =
$$
$$
(c.c.)+i\left(\sqrt{\frac{2\pi}{\omega_i}}\ e^{i(\vec k_i\vec r_1-\omega_i t)}(\vec e_i\cdot\vec\nabla_1) +\sqrt{\frac{2\pi}{\omega_f}}\ e^{-i(\vec k_f\vec r_1-\omega_f t)}(\vec e_f\cdot\vec\nabla_1) \right)\ +
$$
$$
\left(\frac{\pi}{\omega_i}\ e^{2i(\vec k_i\vec r_1-\omega_i t)}+ \frac{\pi}{\omega_f}\ e^{-2i(\vec k_f\vec r_1-\omega_f t)} + \frac{2\pi}{\sqrt{\omega_i\omega_f}}\ (\vec e_i\cdot\vec e_f)\  e^{i[(\vec k_i-\vec k_f)\vec r_1-(\omega_i-\omega_f) t]} \right), \eqno (3.1)
$$
and the same for a positron
$$
V_{int}^{pos}= -i\frac1c\ (\vec A(\vec r_2,t)\cdot\vec\nabla_2) +\frac{1}{2c^2}A^2(\vec r_2,t)\ =
$$
$$
(c.c.) -i\left(\sqrt{\frac{2\pi}{\omega_i}}\ e^{i(\vec k_i\vec r_2-\omega_i t)}(\vec e_i\cdot\vec\nabla_2) +\sqrt{\frac{2\pi}{\omega_f}}\ e^{-i(\vec k_f\vec r_2-\omega_f t)}(\vec e_f\cdot\vec\nabla_2) \right)\ +
$$
$$
\left(\frac{\pi}{\omega_i}\ e^{2i(\vec k_i\vec r_2-\omega_i t)}+ \frac{\pi}{\omega_f}\ e^{-2i(\vec k_f\vec r_2-\omega_f t)} +\frac{2\pi}{\sqrt{\omega_i\omega_f}}\ (\vec e_i\cdot\vec e_f)\  e^{i[(\vec k_i-\vec k_f)\vec r_2-(\omega_i-\omega_f) t]} \right). \eqno (3.2)
$$

Let us consider first the so-called $A^2$  model  and select from
(3) the terms corresponding to Compton scattering, when we have
one incoming photon  and one outgoing photon. In this case, it
follows from the second brackets in (3)
$$
{\tilde V}_{int}(t)=
\frac{2\pi}{\sqrt{\omega_i\omega_f}}(\vec e_i\cdot\vec e_f)\ \left[ e^{i[(\vec k_i-\vec k_f)\vec r_1-(\omega_i-\omega_f) t]} +   e^{i[(\vec k_i-\vec k_f)\vec r_2-(\omega_i-\omega_f) t]}\right]. \eqno (4)
$$

\subsection*{2.1 Schr\"odinger equation}

The evolution of the system can be described by the time-dependent
Schr\"odinger equation:
$$
\left[i\frac{\partial}{\partial t} -H_0 - (V_{int}(t)-{\tilde V}_{int}(t))\right]|\Psi(t)> = {\tilde V}_{int}(t)|\Psi(t)>\eqno(5)
$$
where the total potential of the interaction of positronium with a
photon $V_{int}(t)$  is given by the sum of expressions (3.1) and
(3.2), and the Coulomb potential is included in $H_0$. Let us
express the variables $\vec r_1$ and $\vec r_2$ in terms of the
relative coordinate $\vec{\rho}=\vec r_1-\vec r_2$ and the
coordinate of the positronium center of mass $\vec R = (\vec
r_1+\vec r_2)/2$:
$$
\vec r_1=\vec R+\frac{\vec\rho}{2}, \quad \vec r_2=\vec R-\frac{\vec\rho}{2}. \eqno(6)
$$
Then the conjugate momenta look like
$$
\vec p_1=\frac12\vec p_R+\vec p_\rho, \quad \vec p_2=\frac12\vec p_R-\vec p_\rho. \eqno(7)
$$
This makes it possible to represent all the quantities in $H$ in
terms of the variables $\vec{\rho}$ and $\vec R$ in particular:
$$
H_0=\frac14 p^2_R+ p^2_{\rho}+V(\vec{\rho}), \eqno (8)
$$
and
$$
{\tilde V_{int}}(\vec R, \vec\rho; t)=
\frac{2\pi}{\sqrt{\omega_i\omega_f}}(\vec e_i\cdot\vec e_f)\ e^{i[(\vec k_i-\vec k_f)\vec R-(\omega_i-\omega_f) t]}\left[ e^{i(\vec k_i-\vec k_f)\vec\rho/2} +   e^{-i(\vec k_i-\vec k_f)\vec\rho/2}\right]. \eqno (9)
$$

We denote by $\vec Q = \vec k_i-\vec k_f$ the momentum transferred
from a photon to positronium. In what follows, we need to
calculate the matrix element
$$
{\mathcal M} = \int d^3R\int_{-\infty}^{\infty}dt e^{-i\varepsilon_0 t}<\Psi^-(\vec p_1,\vec p_2,t)|{\tilde V}_{int}(t)|\phi_0>, \eqno (10)
$$
which describes the positronium disintegration  under the action
of the field. Here $\varepsilon_0=-1/8$ denotes the ground
 energy of positronium.

We need to solve Eq. (5) to find the exact wave function for
substituting it in (10), but now we consider the simplest case,
where we neglect the term $(V_{int}-{\tilde V}_{int})$ in this
equation. This term is really small because of the presence of $\omega$ in denominators. Then
$$
<\Psi^-(\vec p_1,\vec p_2,t)|\vec r_1,\vec r_2> = e^{i\vec p_R\vec R - i(E_1+E_2) t}\phi^-(\vec p_\rho,\vec\rho), \eqno (11)
$$
and the function $|\phi^->$ satisfies the hydrogen-like equation
and describes A Coulomb function of the positronium continuum spectrum satisfies the equation:
$$
[p^2_\rho + \triangle_{\rho}-V(\vec\rho)]\phi^-(\vec
p_\rho,\vec\rho)=0, \eqno (12)
$$
In this case, Eq. (10) can be easily integrated, and these integrals give the laws of
conservation of energy and momentum.

\subsection*{2.2 Differential cross section}

The fully differential
cross section (FDCS) for the decay of positronium under the action
of the EM field of the laser is written as
$$
d^3\sigma=\frac{(2\pi)^2\alpha}{\omega_i\omega_f}\sum_{e_i,e_f}|M|^2 \ \times
$$
$$
(2\pi)^4\delta(\omega_i+\varepsilon_0-\omega_f-p^2_1/2 - p^2_2/2)
\delta^3(\vec k_i-\vec k_f-\vec p_1-\vec p_2)\ \frac{d^3 k_f}{(2\pi)^3}\frac{d^3
p_1}{(2\pi)^3}\frac{d^3 p_2}{(2\pi)^3}, \eqno (13)
$$
where the matrix element is given by the expression
$$
M = (\vec e_i\cdot\vec e_f)<\phi(\vec p_\rho)|e^{i\vec Q\cdot\vec\rho/2} +   e^{-i\vec Q\cdot\vec\rho/2}|\phi_0>, \eqno (14)
$$
Here $<\vec\rho|\phi_0>$ and $<\phi(\vec p_\rho)|\vec\rho>$ are
the initial (ground) and final (disintegrated) states of
positronium.  Also $\alpha^2=r_0$ denotes the classical radius of
the electron in atomic units. The sum denotes averaging over the
initial photon polarizations and summing over the final
polarizations. In this case the result does not
depend on the choice of the  polarisation type, either linear or
circular. Approximation (14) will be called the first Born
approximation (FBA) by analogy with the scattering of ordinary
particles.

Further simplification of formula (13) depends on what we actually
measure. In the COLTRIMS detector  only the momenta
of charged particles are measured, but the photon momentum can be
easily calculated from those of  the electron and positron. So, let
us assume that we do not measure the momentum of the positron. Then, after integrating (13) with
respect to $d^3 p_2$, we obtain
$$
d^3\sigma=\frac{(2\pi)^2\alpha}{\omega_i\omega_f}\sum_{e_i,e_f}|M|^2\
(2\pi)\delta(\omega_i+\varepsilon_0-\omega_f-p^2_1/2-(\vec k_i-\vec k_f-\vec p_1)^2/2)\ \frac{d^3
p_1}{(2\pi)^3}\frac{d^3 k_f}{(2\pi)^3}.
\eqno (15)
$$
The phase volume of the photon can be transformed as follows $d^3
k_f=\omega_f^2d\omega_fd\Omega_f/c^3$ with the subsequent
integration with respect to $\omega_f$. In so doing, we assume
that the initial photon frequency is not larger than 5 keV, i.e.
$k_i\sim 1$, which gives some simplification of the integration
results. Finally, we get
$$
\frac{d^3\sigma}{dE_e d\Omega_1d\Omega_f}=FDCS=\frac{\alpha^4}{(2\pi)^3}\ p_1\ \left(1-\frac{p^2_1/2-\varepsilon_0 + (\vec k_i-\vec p_1)^2/2}{\omega_i}\right)\sum_{e_i,e_f} |M|^2. \eqno (16)
$$
For the chosen photon energy, the second term in the brackets
plays the role of a correction. In what follows, we denote
$$
t=\frac{\omega_f}{\omega_i}=\left(1-\frac{p^2_1/2-\varepsilon_0 + (\vec k_i-\vec p_1)^2/2}{\omega_i}\right).
$$
In these notations
$$
Q=\frac{\omega_i}{c}\sqrt{1-2t\cos\theta+t^2},
$$
and $\theta$ is the scattering angle between the vectors $\vec k_i$ and $\vec
k_f$.

We note that
$$
\sum_{e_i,e_f} |M|^2 = \frac12(1+\cos^2\theta)\left|<\phi^-(\vec p_\rho)|e^{i\vec Q\cdot\vec\rho/2}
 +   e^{-i\vec Q\cdot\vec\rho/2}|\phi_0>\right|^2,\eqno (17)
$$
We also note that, after calculating the matrix element,  we must
put $\vec p_\rho= 1/2[\vec p_1-(\vec Q-\vec p_1)]=\vec p_1-\vec
Q/2$ in it, which follows from the law of momentum conservation.

Each term in (17) is analogous to the matrix element for
hydrogen, where $ Z = 1/2 $ should be set. Then (see
\cite{EPJD20})
$$
J_0(\vec Q/2,\vec p_\rho)= \langle\phi^-(\vec p_\rho)|e^{i\vec Q\cdot\vec
\rho/2}|\varphi_0\rangle=
$$
$$
-16\pi\sqrt{\frac{Z^5}{\pi}}
e^{-\pi\zeta/2}\Gamma(1+i\zeta)\frac{[Q^2/4-(p_\rho+iZ)^2]^{-1+i\zeta}}{[(\vec
Q/2-\vec p_\rho)^2+Z^2]^{2+i\zeta}} \  \times
$$
$$
\frac{Q}{2}\left[\frac{Q}{2}-(p_\rho+iZ)\frac{(\vec Q\cdot\vec p_\rho)}{Qp_\rho}\right], \quad  \zeta=-\frac{1}{2p_\rho}, \ Z=\frac 12. \eqno (18).
$$
and
$$
<\phi^-(\vec p_\rho)|e^{i\vec Q\cdot\vec\rho/2} +   e^{-i\vec Q\cdot\vec\rho/2}|\phi_0> =
J_0(\vec Q/2,\vec p_\rho) + J_0(-\vec Q/2,\vec p_\rho).
$$

\section*{3. NUMERICAL CALCULATIONS AND DISCUSSION}.

Thus, we evaluate:
$$
FDCS=\frac{\alpha^4}{(2\pi)^3}p_1 t\sum_{e_i,e_f}|M|^2, \eqno (19)
$$
where
$$
\sum_{e,e_1}|M|^2=\frac12(1+\cos^2\theta)\times
$$
$$
\left|-16\pi\sqrt{\frac{Z^5}{\pi}}\ e^{-\pi\zeta/2}\Gamma(1+i\zeta)\ \frac{Q}{2}\ [(Q^2/4-(|\vec p_1-\vec Q/2| +iZ)^2]^{-1+i\zeta} \ \times\right.
$$
$$
\left.\left[\frac{[Q/2-(|\vec p_1-\vec Q/2|+iZ)\cos\gamma]}{[(\vec p_1-\vec Q)^2+Z^2]^{2+i\zeta}} +
\frac{[Q/2+(|\vec p_1-\vec Q/2|+iZ)\cos\gamma]}{[p_1^2+Z^2]^{2+i\zeta}}\right]\right|^2, \eqno(20)
$$
When calculating the cross section (19), it is possible to use
various systems of units. The cross section is now written in the
atomic units. To get the FDCS in cm $^2/{\rm eV}\cdot {\rm Sr}^2$,
one needs to multiply Eq. (19) by the factor $1.03\cdot 10^{-18}$.
Thus,  $\alpha^4=0.28\cdot 10^{-8} a.u. \to 0.29\cdot 10^{-26}\
{\rm cm}^2/{\rm eV}\cdot {\rm Sr}^2 =0.29\cdot 10^{-2}\ {\rm
barn}/ {\rm eV}\cdot {\rm Sr}^2$.

In (20), the effective charge of positronium is $Z=1/2,\quad \vec
p_\rho =\vec p_1-\vec Q/2$, and also, $$ \cos\gamma=(\vec
Q\cdot\vec p_\rho)/Q p_\rho= \frac{(\vec p_1\cdot\vec
Q)-Q^2/2}{Q\sqrt{p^2_1-(\vec p_1\cdot\vec Q)+Q^2/4}},
$$
$$
(\vec Q\cdot\vec p_1)=k_ip_1[\cos\varphi_1-t\cos\chi]=Qp_1\cos\beta.\eqno (21)
$$
Here
$$
Q=k_i\sqrt{1-2t\cos\theta+t^2}, \quad \cos\chi=\cos\theta\cos\varphi_1+\sin\theta\sin\varphi_1\cos\Phi,
$$
$\varphi_1$ is the angle between $\vec p_1$ and $\vec k_i$, $ \Phi $ is the angle between the planes, in which the vectors
$(\vec k_i, \vec k_f)$ and $(\vec k_i, \vec p_1)$ lie, and $ \theta $
is the scattering angle between the vectors  $\vec k_i$ and $\vec k_f$. We also
repeat that
$$
\zeta = -Z/p_\rho = -1/|2\vec p_1-\vec Q|, \quad t=1-\frac{p^2_1/2-\varepsilon_0 + (\vec k_i-\vec p_1)^2/2}{\omega}.
$$

For reference we present similar expressions for hydrogen from
paper \cite{EPJD20}:
$$
FDCS_H=\frac{\alpha^4}{(2\pi)^3}p_1 t\ \sum_{e_i,e_f}|M_H|^2, \eqno (19.1)
$$
$$
\sum_{e_i,e_f}|M_H|^2=\frac12(1+\cos^2\theta)\times
$$
$$
\left|-16\pi\sqrt{\frac{Z^5}{\pi}}\ e^{-\pi\zeta/2}\Gamma(1+i\zeta)\ Q[Q^2-(p_1+iZ)^2]^{-1+i\zeta} \left[\frac{Q-(p_1+iZ)\cos\chi]}{[(\vec p_1-\vec Q)^2+Z^2]^{2+i\zeta}} \right]\right|^2. \eqno(20.1)
$$
Here in the case of hydrogen
$$
\zeta = -Z/p_1 , \quad t=1-\frac{p^2_1/2-\varepsilon_0}{\omega_i}, \quad Z=1.
$$

\begin{figure}[ht!]
\centering
\includegraphics[scale=0.17]{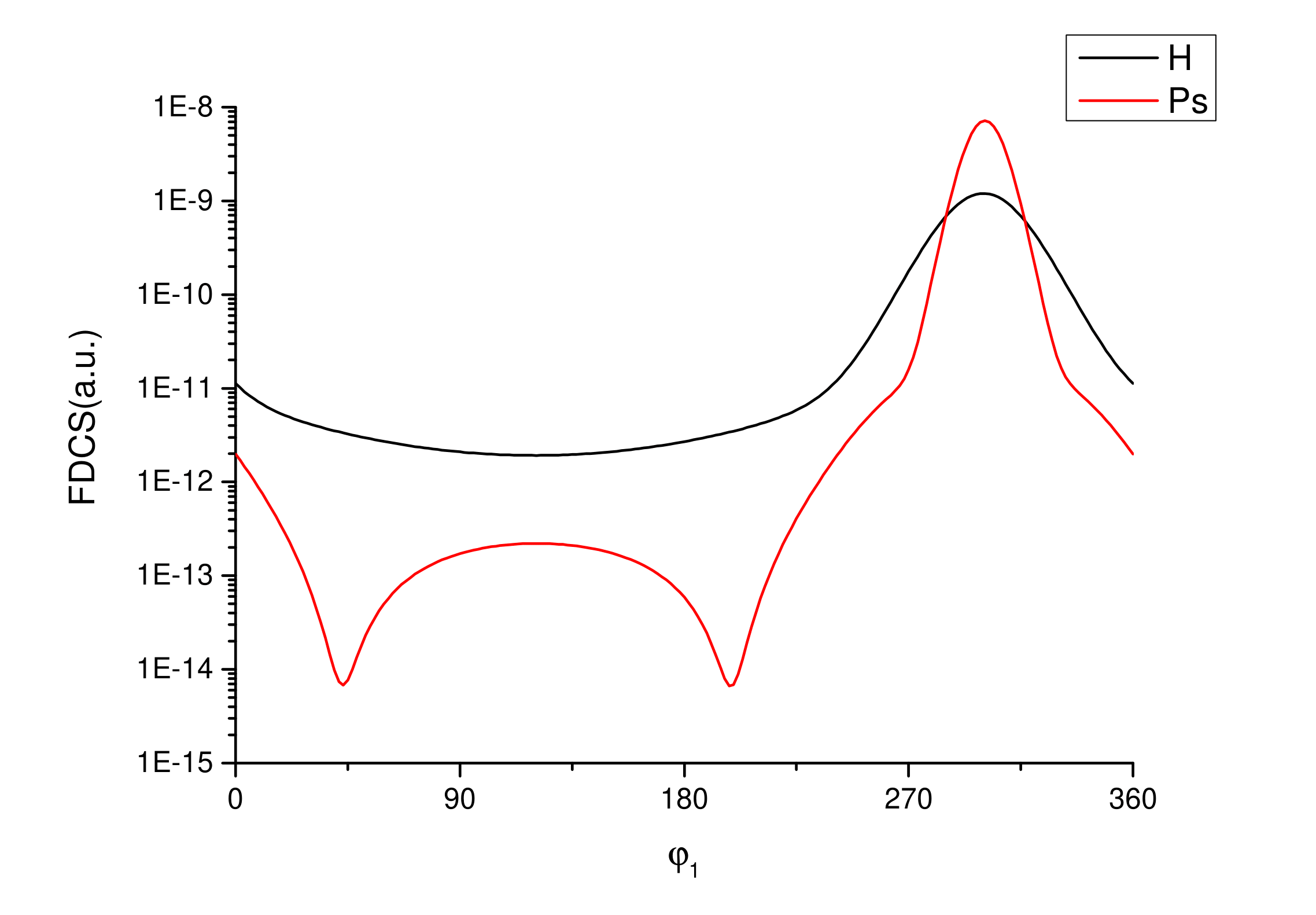} 
\caption{ FDCS (a.u.) as a function of the electron scattering
angle $\varphi_1$ in the case of ionization of the hydrogen atom H
and positronium Ps. $\omega_i=5$ keV, $E_e=27.2$ eV, $\theta=\pi/3$. }
\end{figure}

\begin{figure}[ht!]
\centering
\includegraphics[scale=0.17]{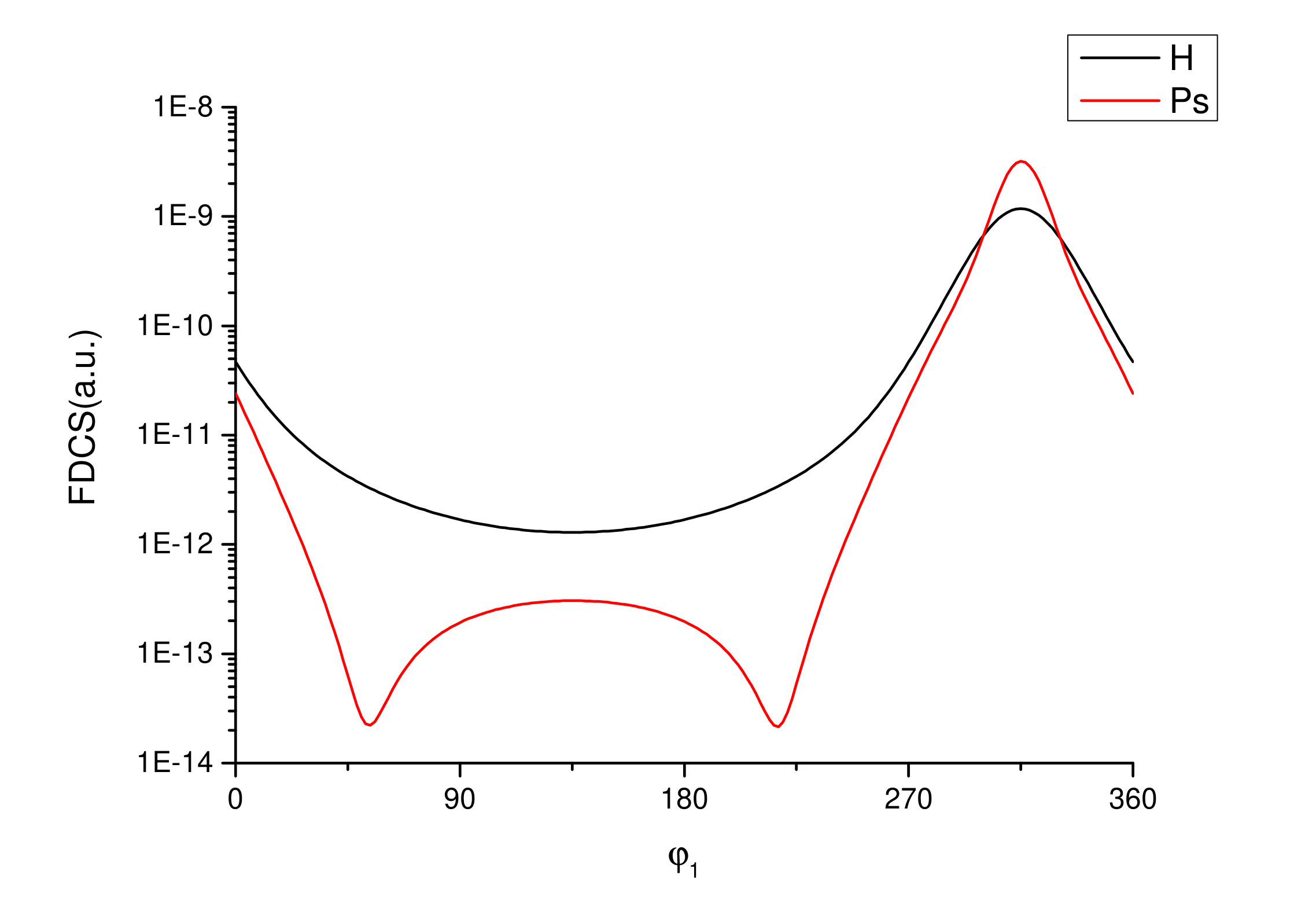}
\caption{ The same as in Fig. 4 for $\theta=\pi/2$.}
\end{figure}

\begin{figure}[ht!]
\centering
\includegraphics[scale=0.17]{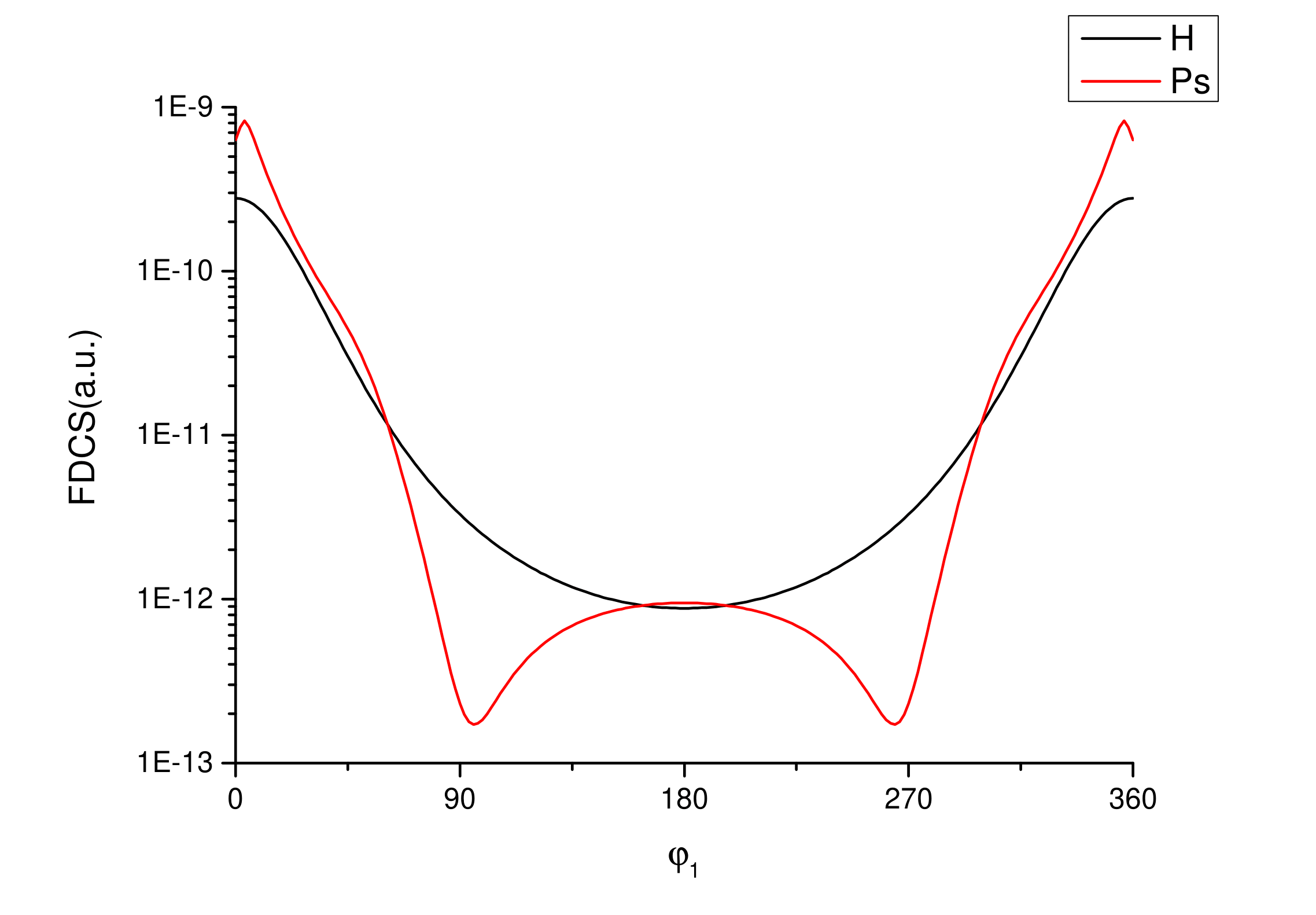}
\caption{ The same as in Fig. 4 for $\theta=\pi$.}
\end{figure}

Let us analyze Figs. 1 - 3, which show the total differential
cross sections of the Compton ionization of the hydrogen atom and
positronium. We should say, that in all calculations we found that the dependence on $t$ is quite weak, so we put everywhere $t=1$. First of all, we see the dominant peaks, when the
angle $\beta$ between the vectors  $\vec p_1$ and $\vec Q$ is
equal to zero. In the theory of particle scattering, this peak is
called  the binary one and corresponds to the
minimum of $|\vec p_1-\vec Q|$. This peak reflects a mechanism,
when the entire transferred momentum is transferred 
to the electron, and the positron remains a "spectator". \
Sufficiently flat second peak of the curve Ps is called the recoil
peak in the theory of particle scattering. In this case, the value
of $|\vec p_1-\vec Q|$ reaches its maximum, and the behavior of
the cross section is determined mainly by the second term in (20),
whose denominator does not depend on the angle $\beta$. In this
region, the positron absorbs a $\gamma$-quantum, and the electron
emits it during the disintegration. The transfer of energy from
the positron to the electron occurs due to the Coulomb interaction
of the particles in positronium. Such a process is possible, but
unlikely, it is reflected in a very large ratio of the magnitudes
of these peaks. In addition, the described process is secondary,
and the electron escapes from the positronium isotropically, which
is reflected in the "flat" character of the backward peak. Thus,
Figs. 1 - 3 present the main mechanisms of the Compton
disintegration of positronium. It is interesting that, at a
sufficiently high photon energy of 5 keV, the backward peak in
hydrogen is not only absent, but turns into a minimum at the same
place, where the Ps curve has the recoil peak. However, the
structural differences between the curves are visible only on a
logarithmic scale.

It should also be noted that the magnitude of the binary peak of
the Ps curve  slightly decreases, while the recoil
peak slightly increases with increasing momentum
transfer $ Q $. This is due to an increase in the value of $|\vec
p_1-\vec Q|$ for a fixed electron energy.

\subsection*{3.1  System of Coulomb resonances}

Let us now consider the behavior of the cross section at the
resonance, where the electron and positron fly  parallel to each
other with the same velocities. This happens, when the Coulomb
number goes to infinity,  $\zeta\to\infty$, or $ \mu =|\vec
p_1-\vec Q/2|\to 0.$  This is possible, when $\vec p_1
\parallel \vec Q$, i.e. in the binary peak region, and for $p_1/k_i=\sin(\theta/2)<1$,
if we set $ t = 1 $ without loss of generality. In this case, it
follows from Eq. (20)
$$
\sum_{e_i,e_f}|M|^2\approx \frac{1}{\mu}\ (1+\cos^2\theta) \frac{32\pi^2Z^5Q^4}{(Q^2/4+Z^2)^6}\exp\left(-\ {\frac{2Z}{Q^2/4+Z^2}}\right). \eqno (22)
$$

As can be seen from (22), a pole appears in the cross section
under the given kinematic conditions, which is characteristic of
infinitely narrow dynamic resonances. However, this pole is not
dangerous when integrating with respect to $d^3p_1$, since we make
a change of variables and compensate this pole with the phase
volume of the integration. However, let us assume that we need to
integrate the FDCS over the photon scattering angle, for example,
to calculate
$$
DDCS_{\varphi_1}= \frac{d^2\sigma}{dE_e d\varphi_1} =2\pi\int_0^\pi\ \sin\theta\ d\theta \int_0^{2\pi}d\Phi\ FDCS. \eqno (23)
$$
This double differential cross section describes the angular and
energy distribution of the electron for any direction of photon
scattering. Let us ask ourselves a question, whether this integral
will converge? To answer this question, we use expression (21) and
write down the expression for $\mu^2$:
$$
\mu^2=p^2_1-p_1Q\left(\cos\varphi_1\sin\frac{\theta}{2}-\sin\varphi_1\cos\frac{\theta}{2}\cos\Phi\right)+\frac{Q^2}{4}. \eqno (24.1)
$$
Recall that for $t=1$, $Q=2k_i\sin(\theta/2)$, and denote $\gamma=
p_1/k$. Omitting intermediate calculations, we rewrite (24.1) in
the form
$$
\mu^2=k_i^2\left[\left(\gamma-\sin\frac{\theta}{2}\right)^2+2\gamma\sin\frac{\theta}{2}\left(1-\cos(\frac{\pi}{2} -\varphi_1-\frac{\theta}{2})\right)\right. \ +
$$
$$
\left.\gamma\sin\varphi_1\sin\theta(1+\cos\Phi)\frac{}{}\right]. \eqno (24.2)
$$
All terms in (24.2) are positive, therefore $\mu=0$ if the
conditions $\gamma=\sin(\theta_0/2), \ (\pi/2
-\varphi_1-\theta_0/2)=0, \ (1+\cos\Phi_0)=0$ are fulfilled
simultaneously. Hence, in particular, it follows that
$\gamma=\cos\varphi_1 < 1$. This is the condition, where one can
expect a singularity in the FDCS. If $p_1>k_i$, or in our case
$E_e>24.4$ eV, the singularities are absent.

Let us put  $\theta=\theta_0+\delta$ and $\Phi=\pi-\lambda$ in
(24.2), where the deviations $\lambda$ and $\delta$ are small.
Then
$$
\mu\approx k\sqrt{\frac14\delta^2+\lambda^2\ \gamma^2(1-\gamma^2)}. \eqno (25)
$$
and the FDCS is completely integrable in the neighborhood of the
points $\theta=\theta_0$ and $\Phi=\pi$, if only $\gamma \neq 1$.
However, in the phase volume of integral (23) in the vicinity of
the point  $\theta=\theta_0$ there is a factor
$\sin\theta_0=2\gamma\sqrt{1-\gamma^2}$, which compensates this
divergence. Thus, integral (23) converges, although numerical
integration requires a certain accuracy in the vicinity of the
zeros of expression (24.2).

\begin{figure}[ht!]
\centering
\includegraphics[scale=0.22]{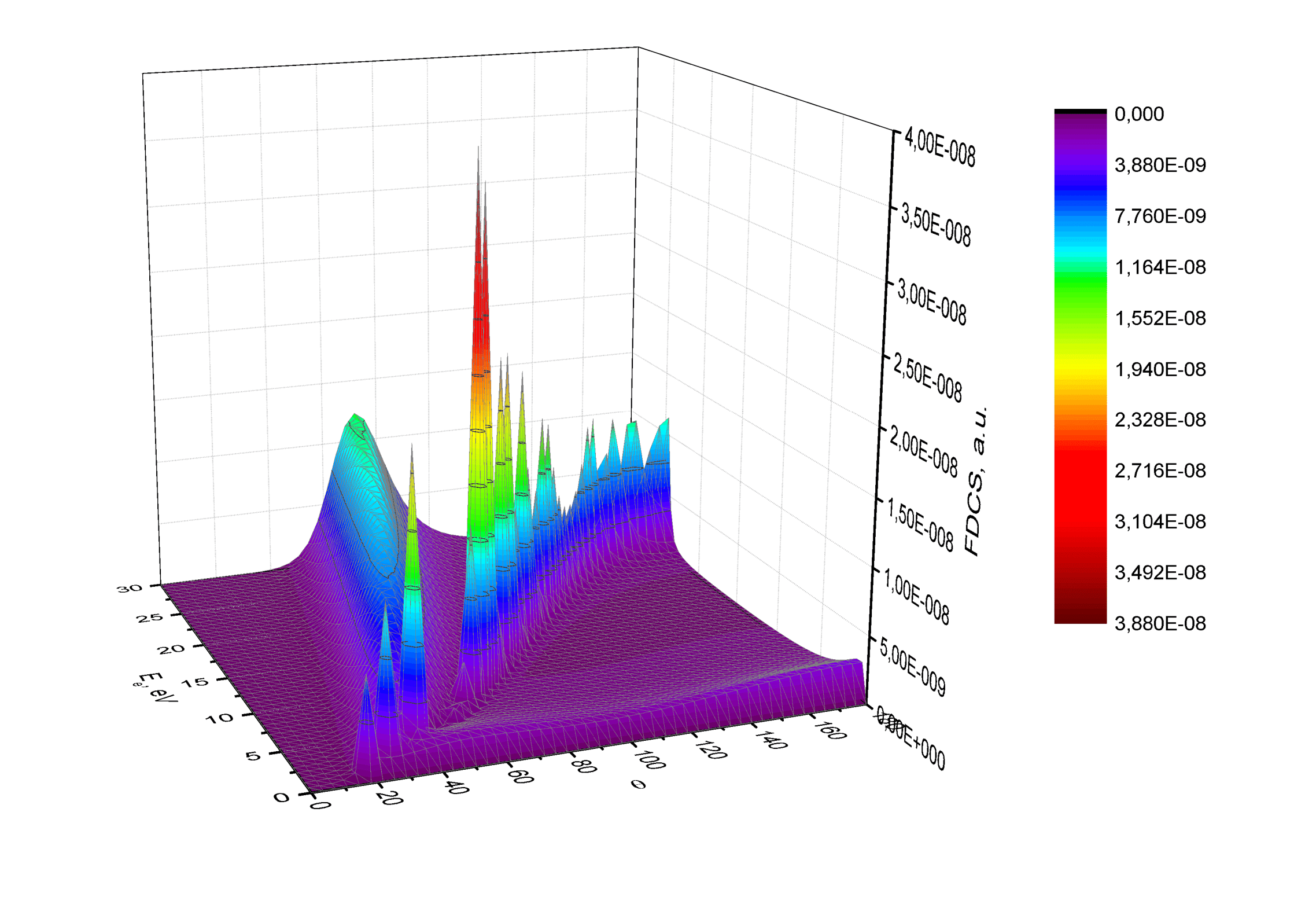}
\caption{ FDCS as a function of the photon scattering angle
$\theta$ and the electron energy $E_e$  under the condition $\vec
p_1 \parallel \vec Q$. The value of $\mu$ is smoothed:
$\mu\to\sqrt{\mu^2+\varepsilon^2},\ \varepsilon=0.01$.  The line
of resonances $\mu\sim 0$ is clearly visible. }
\end{figure}

Fig. 4 shows a 3D plot of the FDCS versus the electron energy and
photon scattering angle in the binary peak region. It should be
said right away that the angle $\beta$  itself depends on the
scattering angles of the photon and the electron according to
(21). Its fixation imposes a constraint on these observed angles.
However, the resonance line is easily seen in the figure. The
calculations were carried out while smoothing the singularity
$\mu\to\sqrt{\mu^2+\varepsilon^2},\ \varepsilon=0.01$.

Let us discuss the above effect. In principle, a similar situation
arises, when a hydrogen atom is ionized by a photon, if the final
state is described by a Coulomb wave. Due to the difference
between the masses of the proton and the electron, the singularity
in the FDCS arises, when the electron energy tends to zero. This
is precisely the energy boundary between the continuum and the
spectrum of bound Coulomb states, which is sometimes
called "the Keldysh swamp". \ In this case, the size of the
system tends to infinity. Formally, we are faced
with an infinite number of thresholds of excitation reactions. As
we can see, the matrix element (20.1) for the hydrogen ionization
has a singularity at $p_1\to 0$. However, in
the case of hydrogen, this pole is compensated by the factor $ p_1
$ in cross section (19.1), and the cross section tends to a constant
as $p_1\to 0$. 

The same situation takes place in positronium, only now the
singularity depends on the angles. And there is a very simple
condition for its occurrence
$E_e=0.5(\omega_i/c)^2\sin^2(\theta/2)$. In this case, the electron
and positron move parallel to each other with the same velocity,
and they can be separated by any distance, including a rather
large one. Formally, such a state can exist for an arbitrarily long time
within the framework of non-relativistic physics, until a second photon is
emitted (even with an energy close to zero), and the
electron-positron pair again falls into a bound state. Thus, a
resonance arises. An interesting question is: how to maintain such
a state for as long as you like?

The described picture resembles the Thomas effect in the classical
description of the capture of an electron by an ion from a target
\cite{Thomas}. That time there was no possibility to
describe the bound state of the final atom in this effect, and it
was described as the parallel motion of an electron and an ion
with the same velocity. From this condition, the trajectory of the
electron motion between heavy fragments was calculated, which gave
a peak in the capture cross section (a second-order process in a
quantum description) .

\vskip 12pt

\subsection*{3.3 Double differential cross section}

Now we consider the double differential cross sections, the
measurements of which are easier. For example, in the case of
cross section (23), we measure only the electron momentum. A 3D
picture of this cross section for positronium is shown in Fig. 5.
Here the photon energy is $\omega_i=3$\ keV. First of all, it is
striking that the cross section is noticeably different from zero
at small angles of electron emission (forward scattering cone of
the photon) and its low energies. The cross section goes to zero
at $E_e\to 0$ in contrast to hydrogen, since there is no
compensating pole here, which we wrote about above. This results
in the sharp peak at low energies. Further, at small angles, two
more peaks of different heights and widths stand out.

 Let us
consider these peaks in more detail in Figs. 6 -- 8, where the same
cross section is presented for the angle $\varphi_1=0$, but at different photon energies. 
These figures also show a similar slice of the cross section for the
hydrogen atom.

Let us pay attention to the middle peak, which becomes discernible
at the photon energy $\omega_i=5$ \ keV, but noticeably grows up as
the photon energy decreases. It is easy to see that this peak
appears at $p_1 = k_i$, or $E_e=0.5(\omega_i/c)^2$. The origin of this
peak has not been fully understood yet. It is not excluded that it
is somehow related to the resonances discussed in the previous
section, although it is difficult to reliably say about this after
integration over the photon scattering angle. Hydrogen has nothing
of this kind.

Let us discuss this peculiarity in the behavior of the cross
section in more detail. In the theory of electron capture from a
target and single ionization of the target by an ion at relatively
low projectile and electron speeds, singularities are observed in
the integral cross sections, when the electron and projectile
velocities coincide. In the theory of capture, this is the above
mentioned  Thomas effect, while in the theory of ionization, this
is the so-called "shoulder" in the single differential cross
section \cite{Schulz}. It is observed quite well at proton
energies of 75 keV and 100 keV. In our case, this is the
coincidence of the electron and positron velocities after the
disintegration of positronium by a photon. Thus, the assumption
about the nature of the discussed peak is quite acceptable.

A wide peak at larger energies is observed for both positronium
and hydrogen. The position of these peaks is also close to each
other. Earlier in the study of hydrogen, it was found that the
FDCS$_H$  has a peak, when the electron takes over the entire
transferred momentum, which corresponds to  Compton scattering at
a free electron. It seems that even after the integration over the
angle of the scattered photon, this trend persists. Indeed, the
peak is seen most clearly, when the photon is scattered into the
backward cone, which leads to forward scattering of the electron.
The integration over all angles smoothes the peak somehow, but
does not override its physics. The presence of the light positron
in positronium slightly changes the position and magnitude of this
peak, but the physical effect remains the same.

\begin{figure}[ht!]
\centering
\includegraphics[scale=0.22]{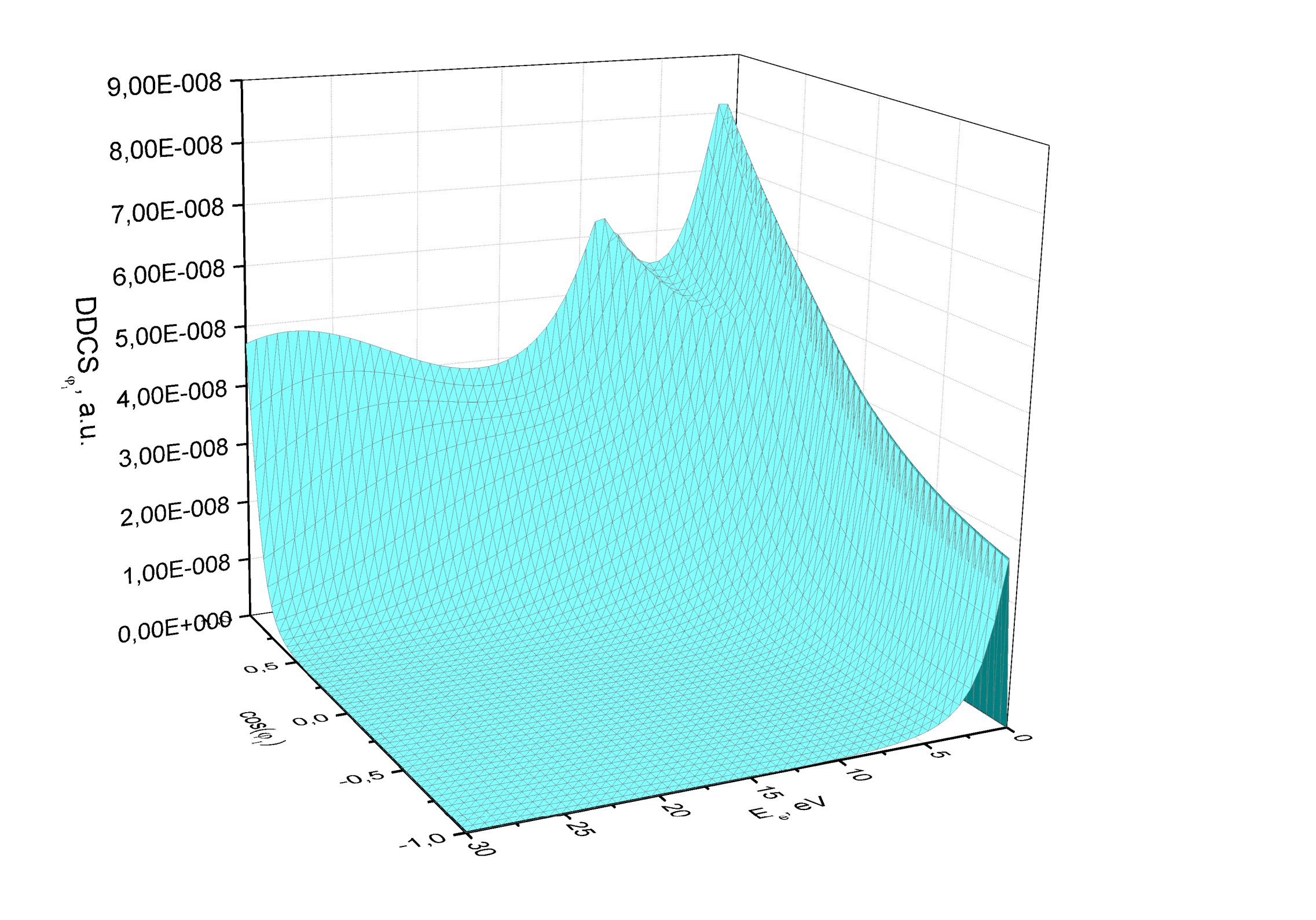}
\caption{ DDCS$_{\varphi_1}$ (23) depending on the electron emission angle
$\varphi_1$ and its energy $ E_e $. The photon energy is $\omega_i=3
\ keV$}
\end{figure}

\begin{figure}[ht!]
\centering
\includegraphics[scale=0.17]{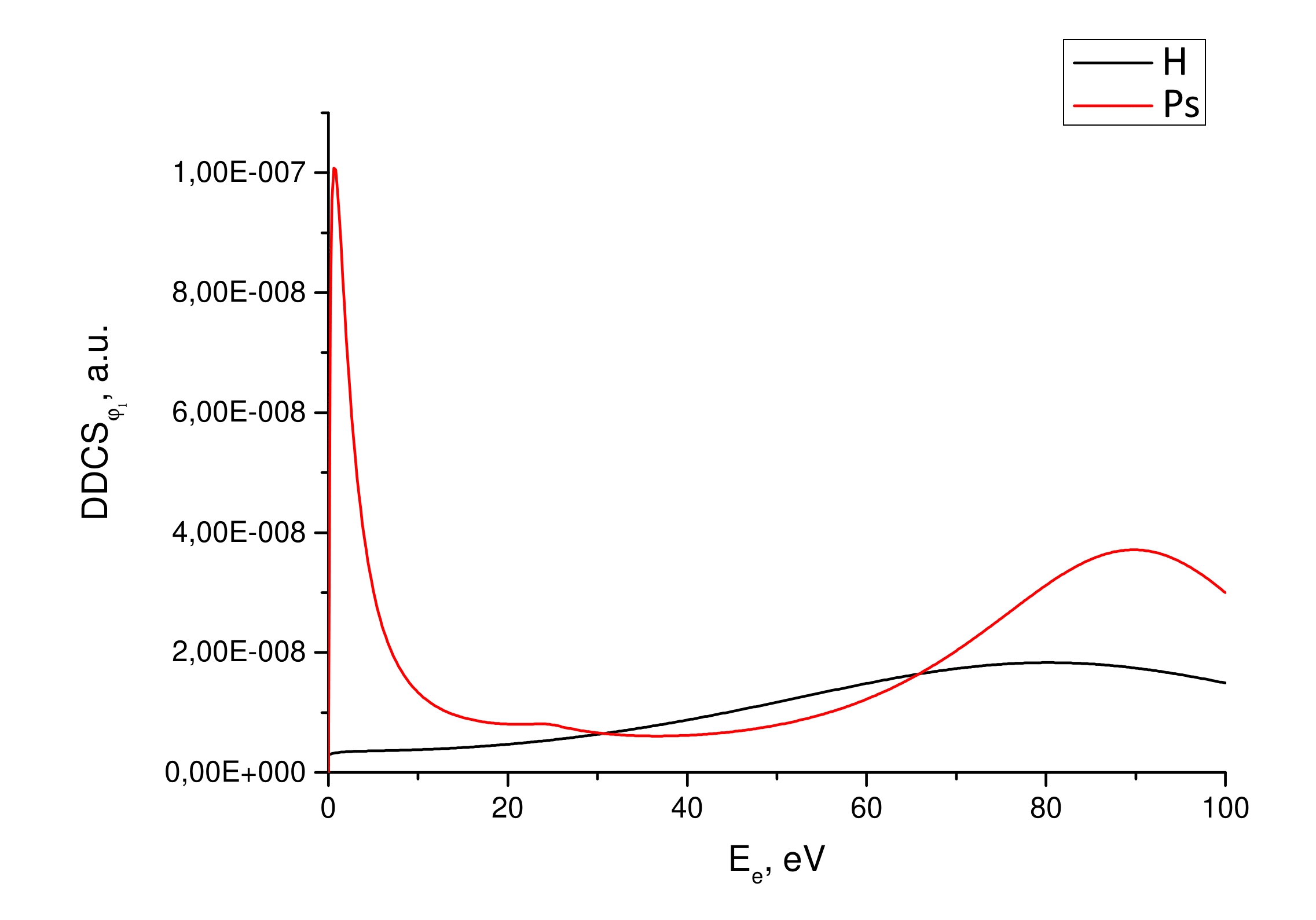}
\caption{ DDCS$_{\varphi_1}$  (23) with the fixed angle $\varphi_1=0$. The
photon energy is $\omega_i=5 \ keV$. The red curve corresponds to
positronium, the black curve corresponds to hydrogen. }
\end{figure}

\begin{figure}[ht!]
\centering
\includegraphics[scale=0.17]{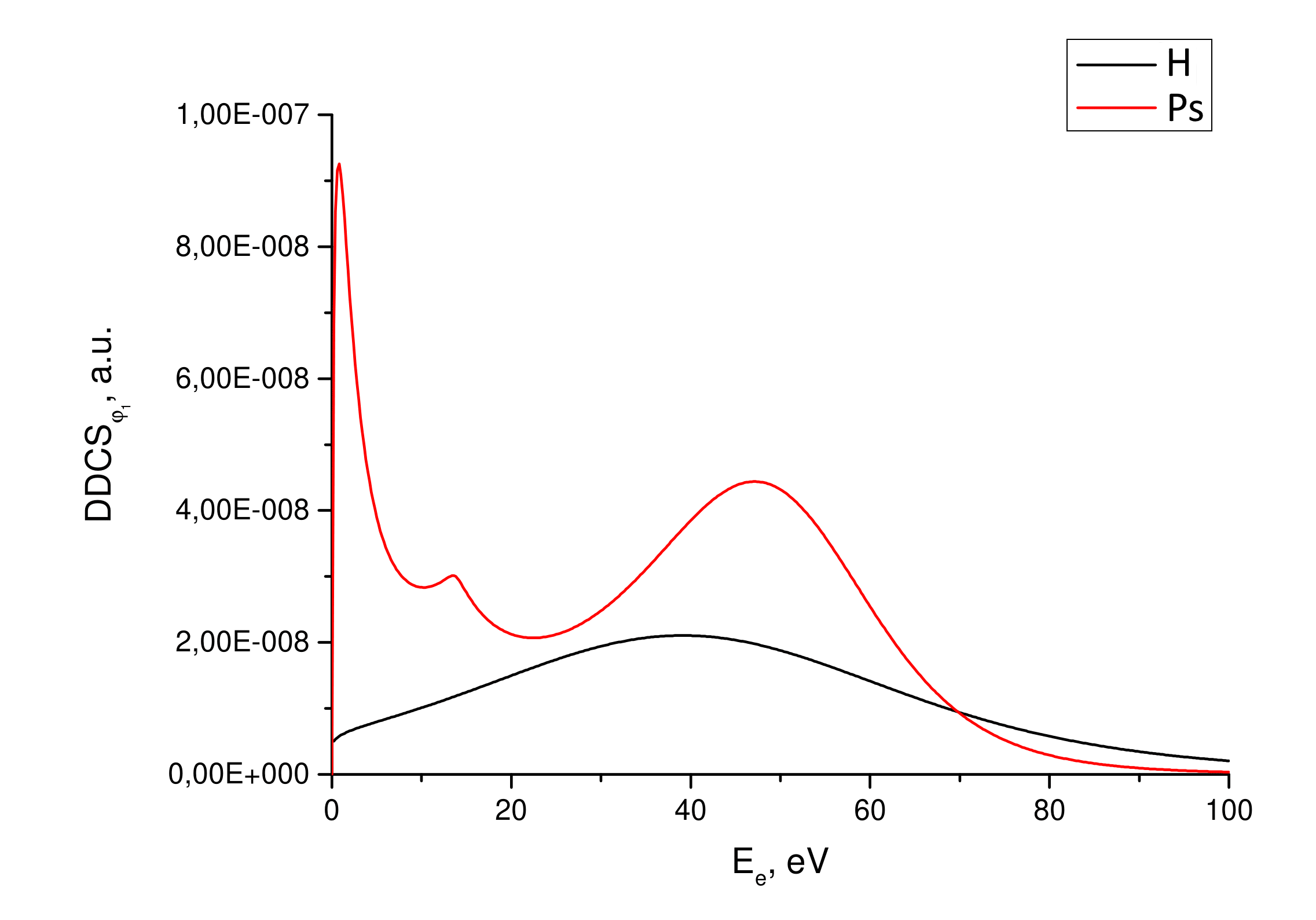}
\caption{ The same as in Fig. 6 for the photon energy $\omega_i=3.75
\ keV$. }
\end{figure}

\begin{figure}[ht!]
\centering
\includegraphics[scale=0.17]{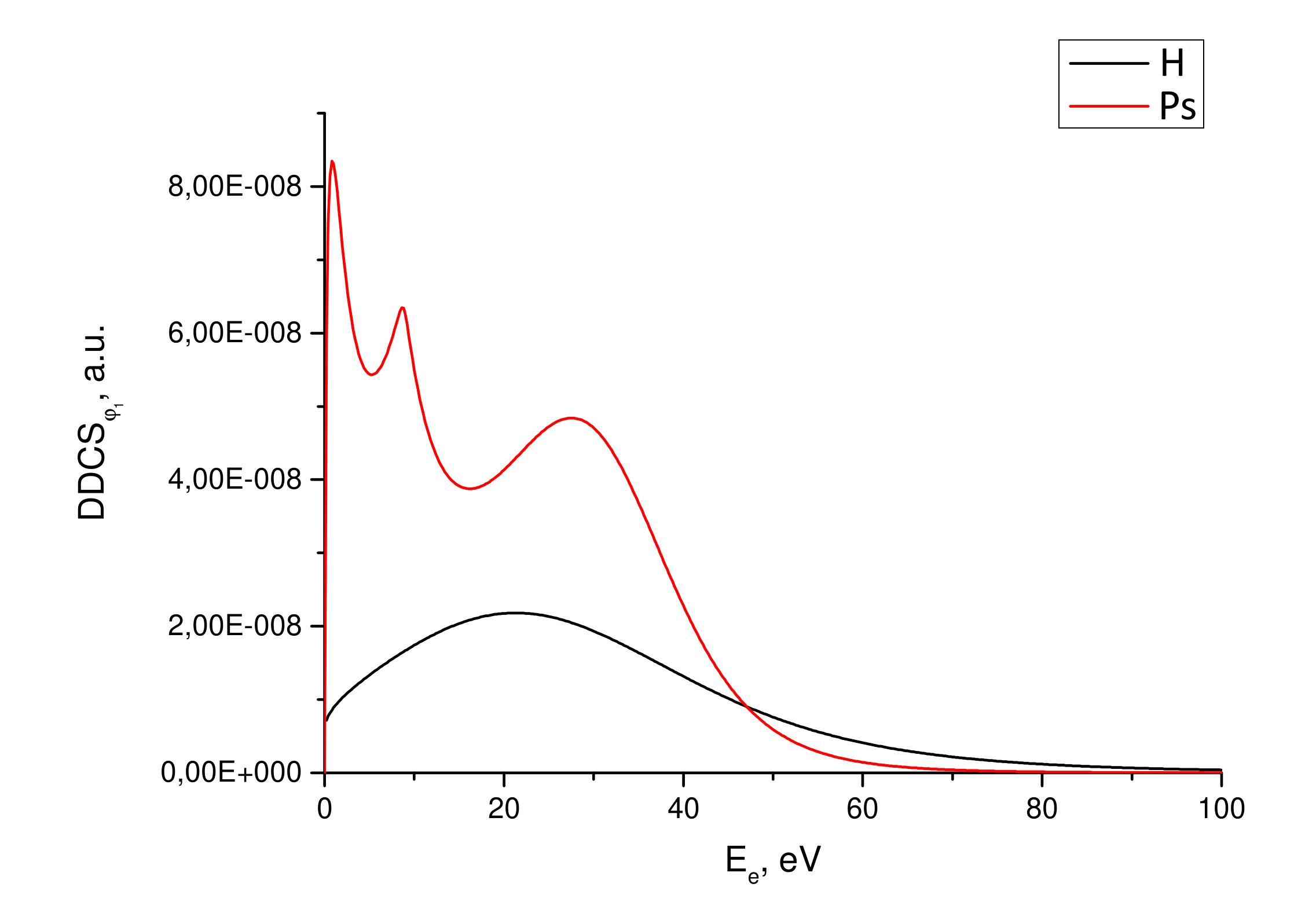}
\caption{The same as in Fig. 6 for the photon energy  $\omega_i=3 \
keV$ .   }
\end{figure}

\section*{4.  CONCLUSION}

We have considered the Compton disintegration of positronium in
the non-relativistic approximation at a photon energy of several
keV. A number of features are noted in the differential cross
sections that distinguish the ionization (disintegration) of
positronium from the ionization of the hydrogen atom. In
particular, an interesting feature of the process are the
resonances that arise in the parallel motion of the electron and
positron with equal velocities after the disintegration. In
hydrogen, only one such resonance state is observed in the total
differential cross section at the energy of the emitted electron
equal to zero. This is due to the use of the Coulomb function as
the final state. For positronium, this point unfolds into a line
of resonances at the electron energies
$E_e=0.5(\omega_i/c)^2\sin^2(\theta/2)$. For the photon
backscattering, the energy of the electron (and the positron),
which is required to provide resonances, reaches its maximum
value, after which the parallel motion of the electron and
positron is no longer possible.

\section*{Acknowledgements}

The authors are grateful to the experimental team headed by  Prof.
R. D\"orner (Institut f\"ur Kernphysik, J. W. Goethe
Universit\"at,  Frankfurt/Main, Germany) whose experiments
inspired us for writing this paper, and for long-time collaboration. 
The work supported in part by the
Heisenberg-Landau program. Yu. P. is grateful to Russian Foundation
for Basic Research (RFBR) for the financial support under grant
No.19-02-00014-a.

\section*{Author contribution statement}

All authors contributed equally to the paper.

\end{document}